\documentclass{webofc}
\usepackage[varg]{txfonts}   % Web of Conferences font
\usepackage{hyperref}
\usepackage{url}
\usepackage[normalem]{ulem}
\hypersetup{colorlinks=true,citecolor=blue,urlcolor=blue,linkcolor=blue}
\begin{document}
\title{Quarkonium suppression in strongly coupled plasmas}

\author{\firstname{Bruno} \lastname{Scheihing-Hitschfeld}\inst{1}\fnsep\thanks{\email{bscheihi@kitp.ucsb.edu}, speaker at HP2024} \and
        \firstname{Xiaojun} \lastname{Yao}\inst{2}\fnsep\thanks{\email{xjyao@uw.edu}}
        % etc.
}

\institute{Kavli Institute for Theoretical Physics, University of California, Santa Barbara, California 93106, USA 
\and
           InQubator for Quantum Simulation, Department of Physics, University of Washington, Seattle, WA 98195, USA
          }

\fnsep{\!\!\!\!\!\!\!\!IQuS@UW-21-100}

\abstract{Suppression of open heavy quarks and quarkonia in heavy-ion collisions are among the most informative probes of quark-gluon plasma (QGP). Interpreting the full wealth of data obtained from the collision events requires a precise theoretical understanding of the evolution of heavy quarks and quarkonia as they propagate through a strongly coupled plasma.
Such calculations require the evaluation of a gauge-invariant correlator of chromoelectric fields. This chromoelectric correlator encodes all the characteristics of QGP that the dissociation and recombination dynamics of quarkonium are sensitive to, which is to say can in principle measure. We review its distinctive qualitative features at weak coupling in QCD up to next-to-leading order and at strong coupling in $\mathcal{N}=4$ SYM using the AdS/CFT correspondence, as well as its formulation in Euclidean QCD.
Furthermore, we report on recent progress in applying our results to the calculation of the final quarkonium abundances after propagating through a cooling droplet of QGP, which illustrates how we may learn about QGP from quarkonium measurements. We devote special attention to how the presence of a strongly coupled plasma modifies the transport description of quarkonium, in comparison to approaches that rely on weak coupling approximations to describe quarkonium dissociation and recombination.
}
\maketitle

\section{Introduction: What properties of QGP does quarkonium probe?}

The dynamics of quarkonium in quark-gluon plasma (QGP) produced in heavy-ion collisions is a topic of high interest, as it has long been recognized~\cite{Matsui:1986dk,Karsch:1987pv} that it can serve as a direct probe of the confinement-deconfinement transition. Indeed, suppression of quarkonia is often referred to as a signature of the formation of deconfined QCD matter.

Recent studies that combine open quantum systems and effective field theory techniques have achieved a systematic field-theoretical framework for heavy quark-antiquark pairs ($Q\bar{Q}$) of small separation. Due to the large heavy quark mass $M$, bound states of the $Q\bar{Q}$ pair can be described in the non-relativistic approximation $v \ll 1$. Furthermore, the distance separation $r$ between the $Q\bar{Q}$ pair is smaller than the relevant medium scales such as the inverse of temperature $1/T$. These two conditions imply that the time evolution of the $Q\bar{Q}$ pair can be described by potential nonrelativistic QCD (pNRQCD)~\cite{Brambilla:1999xf}, which is constructed by systematic expansions in $1/M$ and $r T$. The degrees of freedom of this effective theory are the wavefunctions of the $Q\bar{Q}$ pairs, which can be either in a color singlet or in a color octet state, together with the light QCD degrees of freedom that constitute the thermal environment. At leading order in this EFT, quarkonium dissociation and recombination are described as singlet-octet transitions because $Q\bar{Q}$ bound states are color singlet states at this order.
These transitions between color singlet and color octet states are induced by the QGP medium via a color dipole interaction. One may thus follow the dynamics of quarkonium by tracking the reduced density matrix
\begin{align}
    \rho_{Q\bar{Q}}(t) = {\rm Tr}_{\rm QGP} \left[ U(t) \rho_{\rm tot}(t=0) U^\dagger(t) \right] \, , \label{eq:oqs-general}
\end{align}
where $\rho_{\rm tot}(t=0)$ is the initial density matrix of the whole system.

This EFT framework allows one to formulate the gauge-invariant correlation functions that one needs to calculate to describe in-medium quarkonium dynamics, or conversely, that can be extracted or constrained from quarkonium suppression data. We refer to these as \textit{Generalized Gluon Distributions} (GGDs)~\cite{Nijs:2023dbc}, which have been studied in QCD up to NLO~\cite{Binder:2021otw,Scheihing-Hitschfeld:2022xqx} and in $\mathcal{N}=4$ SYM in the strong coupling limit~\cite{Nijs:2023dks}.
Explicitly, they are given by
\begin{align} 
\label{eq:g++-definition}
[g_{\rm adj}^{++}]^>(t) &\equiv \frac{g^2 T_F }{3 N_c}  \big\langle E_i^a(t)W^{ac}(t,+\infty) 
W^{cb}(+\infty,0) E_i^b(0) \big\rangle_T \, , \\ \label{eq:g---definition}
[g_{\rm adj}^{--}]^>(t) &\equiv \frac{g^2 T_F }{3 N_c} \big\langle W^{dc}(-i\beta - \infty, -\infty)
W^{cb}(-\infty,t)  E_i^b(t)
E_i^a(0)W^{ad}(0,-\infty)  \big\rangle_T  \, .
\end{align}
They have also been connected to correlation functions that are calculable in Euclidean QCD~\cite{Scheihing-Hitschfeld:2023tuz}, paving the way for a Lattice QCD determination of in-medium quarkonium transport properties.

While GGDs have been formulated and calculated, 
their application to phenomenology so far has relied on further simplifying assumptions. For example, the Quantum Brownian Motion limit~\cite{Brambilla:2016wgg-2017zei} further assumes that the temperature of the surrounding QGP is large compared to the energy level spacings of quarkonium, i.e., $T \gg |\Delta E|$, thus only being sensitive to the low frequency limit of the GGDs. In this limit, the dynamics of quarkonium in medium takes the form of a Lindblad equation $\frac{d\rho}{dt} = \mathcal{L}[\rho(t)]$.

Another example is the Quantum Optical Limit~\cite{Yao:2018nmy}, which assumes a semiclassical description of the dynamics of quarkonium, and neglects the evolution of the off-diagonal elements of the density matrix of the heavy quark pair. The dynamics of quarkonium take the form of a Boltzmann equation
$\frac{Df}{dt} = \mathcal{C}[f(t)]$, where the GGDs enter in the ``collision integral'' evaluated at the energy transfer required to go from one state to the other. Concretely, if written in terms of $[g_{\rm adj}^{++}]^>(\omega)$, the relevant frequency for each transition is $\omega \sim - |\Delta E|$.
While this formulation is able to describe situations where the temperature is of the order of the energy level spacings $T \sim |\Delta E|$, it makes approximations in neglecting quantum aspects and in neglecting QGP correlations in time, over which it is hard to justify one has quantitative control, especially at strong coupling.

These approaches have had a good measure of success in describing quarkonium suppression data. However, the approximations used in either of these approaches weaken the extent to which it is possible to use data to gain deeper understanding of the underlying dynamics of QCD. In what follows, we discuss the qualitative novelties that we have found by carrying out explicit calculations of the GGDs, and lay out the challenges that need to be met in order to describe quarkonium in QGP from first principles.

\section{Lessons from a strongly coupled plasma}

Given that the GGDs characterize the dynamics of quarkonium in medium, we can gain insight about the dynamics  by simply looking at the result of calculations of the GGDs in different scenarios.

\begin{figure}
    \centering
    \includegraphics[width=0.75\textwidth]{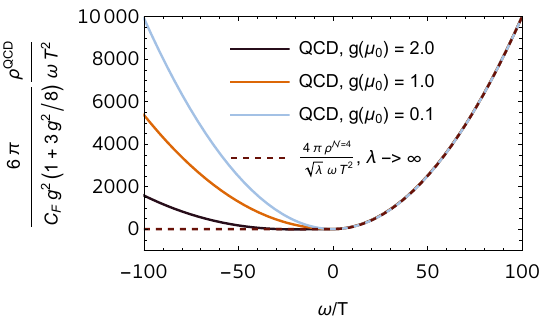}
    \caption{Spectral function for quarkonium transport in weakly coupled QCD with 2 light massless quarks for different values of the coupling at the reference scale $\mu_0 \approx 8.1 T$ (solid lines) and strongly coupled $\mathcal{N}=4$ SYM (dashed line). The QCD coupling constant is evolved to high energies using the 2-loop QCD beta function. The curves are normalized such that their $\omega \gg T$ behavior agree. As such, this plot is illustrative of differences at negative frequency $\omega$.}
    \label{fig:spectral-2sided}
\end{figure}

Figure~\ref{fig:spectral-2sided}, extracted from~\cite{Nijs:2023dbc}, shows a comparison between the spectral function $\rho_{\rm adj}^{++}$ associated to the GGDs (related by $\rho_{\rm adj}^{++}(\omega) = (1 - e^{-\beta \omega} ) [g_{\rm adj}^{++}]^>(\omega) $) in two theory calculations: a weakly coupled NLO calculation in QCD~\cite{Binder:2021otw}, for selected values of the coupling strength, and a strongly coupled calculation in $\mathcal{N}=4$ SYM~\cite{Nijs:2023dks}. When $\omega \gg T$, all of the theory calculations give a result proportional to $\omega^3$. We therefore use this as a reference in order to normalize the curves and compare the qualitative features of the GGDs at negative frequencies, which is the regime of interest for the quantum optical limit (we will discuss the Brownian motion limit momentarily). As we can see from the figure, the main feature as a function of the coupling is that the negative frequency side gets suppressed in amplitude relative to the positive frequency side as the coupling is increased, showing a compatible picture between the QCD and $\mathcal{N}=4$ results. The plot is extended to $|\omega|/T = 100$ to show that this behavior is visible in the regime where perturbation theory is reliable ($|\omega| \gg T, \Lambda_{\rm QCD}$).

The situation at low frequency is qualitatively different: on the one hand, perturbation theory gives a nonzero limit of $\rho_{\rm adj}^{++}(\omega)/\omega$ as $\omega \to 0$, which matches the analog calculation for the heavy quark diffusion coefficient at $\mathcal{O}(g^4)$. On the other hand, the strongly coupled $\mathcal{N}=4$ calculation yields a vanishing low-frequency limit $\lim_{\omega \to 0} \rho_{\rm adj}^{++}(\omega)/\omega = 0$, in stark contrast with the nonzero value of the heavy quark diffusion coefficient in strongly coupled $\mathcal{N}=4$ SYM~\cite{Casalderrey-Solana:2006fio}.

Put together, these observations may have profound consequences for quarkonium transport in the strongly coupled medium that is QGP formed in heavy ion collisions. If a non-perturbative calculation of the GGDs in QCD at physical values of the coupling shows the same features as the $\mathcal{N}=4$ SYM result we just described, then both the quantum optical and quantum Brownian motion limits would yield trivial dynamics.

However, one should not conclude from here that quarkonium undergoes trivial dynamics. Rather, the situation in both approaches can be understood as having chosen an expansion parameter, in this case $rT \ll 1$, and having used this hierarchy of scales to argue that the terms that are kept in each approach are parametrically larger than the neglected terms, with some $\mathcal{O}(1)$ coefficient in front to be determined by calculations. What the strongly coupled $\mathcal{N}=4$ calculation shows is that this $\mathcal{O}(1)$ coefficient vanishes in this theory, rendering ineffective what seemed to be a perfectly good expansion.

To be precise, the expansion that breaks down in this situation is \textit{not} the power counting in powers of $rT$, but rather the Markov approximation that both the quantum optical and quantum Brownian motion limits make based on this hierarchy. One is therefore left with dynamical equations of motion for the quarkonium density matrix that have a perturbatively controlled $rT$ expansion, but in which the leading contributions are driven by non-Markovian effects, i.e., memory terms coming from finite time QGP correlations.

Therefore, the challenge for theory efforts is to develop transport formalisms in which these memory effects can be accounted for, so that the fundamental QCD properties encoded in the GGDs may be probed and extracted from experimental data.

\section{Comparing weakly and strongly coupled pictures}

As a first step towards characterizing the kind of phenomenology that non-Markovian dynamics can lead to, we now compare the total $\Upsilon(1S)$ formation probability from initially unbound color octet states by evolving the $Q\bar{Q}$ density matrix in time-dependent perturbation theory using the different GGD calculations that are currently available. The small parameter $r T$ allows one to expand the time evolution operators in Eq.~\eqref{eq:oqs-general} and keep only terms at the first nontrivial order, leading to
\begin{align}
    \langle nl | \, \rho_{Q \bar{Q}}(\tau_f) \, | nl \rangle = \!\! \int_{\tau_i}^{\tau_f} \!\!\!\!\! d\tau_1 \!\! \int_{\tau_i}^{\tau_f} \!\!\!\!\! d\tau_2 \, [g_{\rm adj}^{--}]^{>}({\tau_2,\tau_1}) \, \langle nl | U^{\rm singlet}_{[\tau_f,\tau_1]} r_i U^{\rm octet}_{[\tau_1, \tau_i]} | \psi_0 \rangle \big( \langle nl | U^{\rm singlet}_{[\tau_f,\tau_2]} r_i U^{\rm octet}_{[\tau_2, \tau_i]} | \psi_0 \rangle \big)^\dagger \, , \nonumber
\end{align}
where $| \psi_0 \rangle$ is the initial wavefunction for the relative position coordinate of the $Q\bar{Q}$ pair in the octet state. $U^{\rm singlet}_{[t,t']}$, $U^{\rm octet}_{[t,t']}$ are, respectively, the time evolution operators from time $t'$ to time $t$ for states in the singlet and octet representations in the absence of transitions.

\begin{figure}[t]
    \centering
    \includegraphics[width=0.75\textwidth]{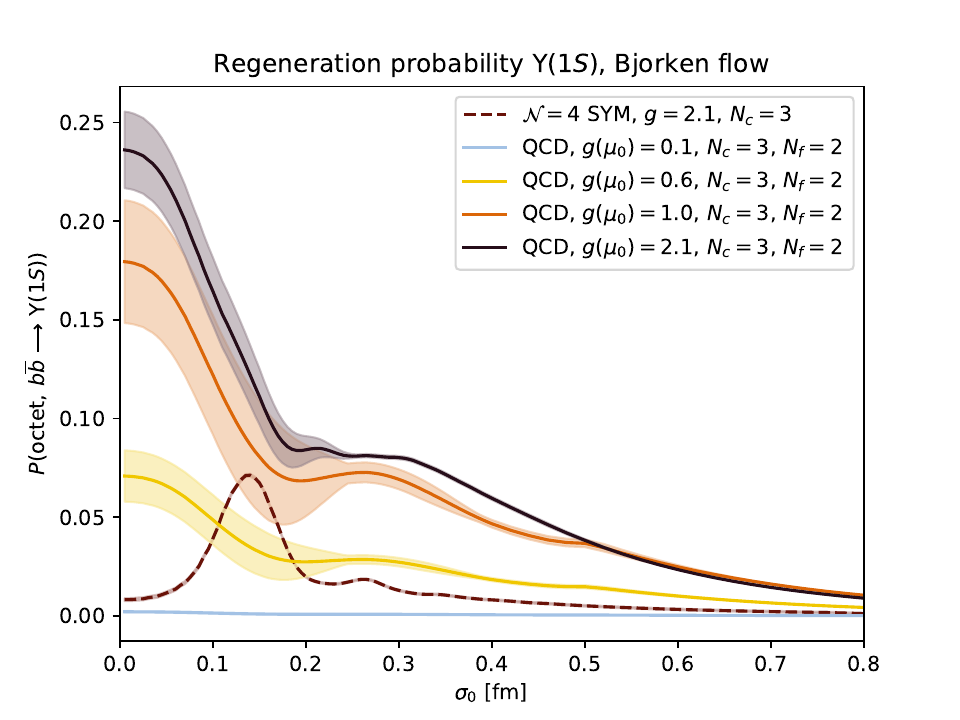}
    \caption{Regeneration/formation probability for an $\Upsilon(1S)$ state as a function of the initial separation $\sigma_0$ between the two heavy quarks, for GGDs with different values of the coupling in QCD at NLO. The temperature profile experienced by the heavy quark pair was set to be given by Bjorken flow scaling, $T(\tau) = (\tau_f/\tau)^{1/3} T_f $, with $T_f = 155 \, {\rm MeV}$, $\tau_i = 0.6 \, {\rm fm/c}$, and $\tau_f = 10 \, {\rm fm/c}$. The initial condition for the wavefunction in the radial component of the relative coordinate was given by $\psi_0(r) \propto r Y_{1m}(\theta, \varphi) \exp(- r^2/(2\sigma_0^2)) $, where $Y_{1m}$ is a spherical harmonic. The reason to choose $\ell=1$ as the initial state is that its transition to the $1S$ state is allowed by the dipole interaction of pNRQCD at the order we work in the EFT. The wavefunction is appropriately normalized to have unit probability, and the final result is averaged over $m$. The color bands quantify the error associated to truncation effects when solving for the evolution of the quarkonium state. Figure reproduced from~\cite{ScheihingHitschfeld:2024fmf}.}
    \label{fig:prob}
\end{figure}

To illustrate this formula, we plot it as a function of different initial conditions in Figure~\ref{fig:prob}, parametrized by different spatial widths of the octet wavepacket $\sigma_0$. To do this, we need to specify an interaction potential model to construct the time evolution operators, which we take to be a Karsch-Mehr-Satz potential~\cite{kms:1998} for the singlet, and no potential for the octet. 
After specifying the model and the temperature of the system as a function of time, the only remaining parameter that needs to be specified in this description is the coupling constant, which for present purposes we set to $\lambda = 13$, corresponding to $g \approx 2.1$.

We see from Figure~\ref{fig:prob} that the QCD NLO correlator and the strongly coupled $\mathcal{N}=4$ correlator exhibit different features. The correlators calculated in a weak coupling expansion yield a generally decreasing regeneration probability as a function of the initial heavy quark separation, with a plateau beginning at around $\sigma_0 \sim 0.15$ fm, which can be identified as the size of the $\Upsilon(1S)$ state, and falling off again when the initial separation $\sigma_0$ is too large for a bound state to be efficiently formed. On the other hand, the regeneration probability obtained with the strongly coupled correlator exhibits peak structures around values of $\sigma_0$ close to the size of bottomonium bound states, with the dominant peak being indicative of the final $1S$ state having the biggest overlap with the initial state with $\sigma_0 \sim 0.13$ fm, but also with indications of having some overlap with the in-medium $2S$ state and even the $3S$ state.

This qualitative difference between the (re)generation probabilities with different correlators can be understood in terms of the trend we observed in Figure~\ref{fig:spectral-2sided}. Namely, the vanishing of the spectral function at negative frequencies in the strongly coupled case suppresses the total regeneration probability because the microscopic processes that would lead to Markovian terms in the density matrix evolution equation are absent. The only source of enhancement in the probability is a large wavefunction overlap. Conversely, the GGDs calculated in perturbation theory lead to a larger regeneration probability because such microscopic processes are present (that is to say, the spectral function doesn't vanish at negative frequencies). One way of interpreting this is in terms of the absence/presence of quasiparticles in the medium that can form ``asymptotic'' states with which quarkonium can scatter: a weakly coupled medium will have such quasiparticles and thus a large phase space to emit (gluon) radiation, but a strongly coupled medium need not. 

It is interesting to note that the change between $g = 0.6$ and $g = 1$ in Figure~\ref{fig:prob} is much larger than the change between $g = 1$ and $g = 2.1$, even though the latter is a jump by a bigger relative factor. We interpret as this being due to a compensation of two effects: on the one hand, a larger coupling enhances the regeneration probability, but on the other, the jump between $g=1$ and $g = 2.1$ makes the spectral function $\rho_{\rm adj}^{++}(\omega)$ much closer to zero at negative frequencies. In turn, this can be taken as indicative of the increasing relative importance of non-Markovian effects as the coupling is increased.

\section{Outlook: Non-Markovian quantum transport and what we will learn from lattice QCD calculations}

The generalized gluon distributions provide a unique window to have a glimpse at finite-frequency properties of QGP with experimental measurements of quarkonium suppression. However, we have found that in order to reliably extract this information from experimental data, it is necessary to improve the currently available transport formalisms to the point where non-Markovian effects can be reliably accounted for. This is crucial for the interpretability of phenomenological models in terms of QCD quantities, because we do not know whether the non-perturbative GGDs in QCD are close enough to the $\mathcal{N}=4$ result to decide whether non-Markovian effects are quantitatively subdominant or not in heavy-ion collisions.

This is why a lattice QCD calculation of the Euclidean version of the GGDs will be very informative. As we have shown in~\cite{Scheihing-Hitschfeld:2023tuz}, the Euclidean correlator
\begin{equation}
	G_{\rm adj}(\tau) = \frac{g^2T_F }{3 N_c} \big \langle E_i^a(\tau) \mathcal{W}^{ab}(\tau,0) E_i^b(0) \big\rangle_T \, , \label{eq:euclidean-G}
\end{equation}
is the analytic continuation of $[g_{\rm adj}^{++}]^>(t)$ to Euclidean time. While carrying out the analytic continuation from lattice QCD data calculated in imaginary time back to real time is a formidable problem, qualitative insights into quarkonium physics may be obtained directly from Euclidean data. 

The reason why such insight might be obtained is the following: the relevance of non-Markovian effects may be quantified by the degree to which the usual time reversal symmetry of the spectral function $\rho_{\rm adj}^{++}(\omega) = - \rho_{\rm adj}^{++}(-\omega)$ is broken. In a sense, the $\mathcal{N}=4$ result maximally breaks this symmetry because the even and odd parts of $\rho_{\rm adj}^{++}(\omega)$ have the same magnitude. Crucially, we showed in~\cite{Scheihing-Hitschfeld:2023tuz} that breaking this symmetry is equivalent to breaking the $\tau \to \beta - \tau$ symmetry that would be present in Eq.~\eqref{eq:euclidean-G} if the Wilson line were absent. One can therefore use the degree to which this symmetry is broken in $G_{\rm adj}(\tau)$ as an indication of how relevant non-Markovian effects are for in-medium quarkonium dynamics in QCD.

{\small The work of BSH is supported by grant NSF PHY-2309135 to the Kavli Institute for Theoretical Physics (KITP), and by grant 994312 from the Simons Foundation. XY is supported by the U.S. Department of Energy, Office of Science, Office of Nuclear Physics, InQubator for Quantum Simulation (IQuS) (https://iqus.uw.edu) under Award Number DOE (NP) Award DE-SC0020970 via the program on Quantum Horizons: QIS Research and Innovation for Nuclear Science.}

\end{document}